# Extreme precipitation events in the Mediterranean: Spatiotemporal characteristics and connection to large-scale atmospheric flow patterns

Nikolaos Mastrantonas (1, 2), Pedro Herrera-Lormendez (1), Linus Magnusson (2), Florian Pappenberger (2), and Jörg Matschullat (1)

(1) Technische Universität Bergakademie Freiberg (TUBAF), Germany; (2) European Centre for Medium-Range Weather Forecast (ECMWF), United Kingdom

**Corresponding author**: Nikolaos.Mastrantonas@doktorand.tu-freiberg.de

**Abstract**

The Mediterranean region is strongly affected by Extreme Precipitation Events (EPEs), sometimes leading to severe negative impacts on society, economy, and the environment. Understanding such natural hazards and their drivers is essential to mitigate related risks.

Here, EPEs over the Mediterranean between 1979 and 2019 are analyzed, using ERA5, the latest reanalysis dataset from ECMWF. EPEs are determined based on the 99$^{th}$ percentile of the daily distribution (P99). The different EPE characteristics are assessed, based on seasonality and spatiotemporal dependencies. To better understand the connection to large-scale atmospheric flow patterns, Empirical Orthogonal Function (EOF) analysis and subsequent non-hierarchical K-means clustering are used to quantify the importance of weather regimes to EPE frequency. The analysis is performed for three different variables, depicting atmospheric variability in the lower and middle troposphere: Sea level pressure (SLP), temperature at 850 hPa (T850), and geopotential height at 500 hPa (Z500).

Results show a clear spatial division in EPEs occurrence, with winter and autumn being the seasons of highest EPEs frequency for the eastern and western Mediterranean, respectively. There is a high degree of temporal dependencies with 20% of the EPEs (median value based on all studied grid-cells), occurring up to 1 week after a preceding P99 event at the same location. Local orography is a key modulator of the spatiotemporal connections and substantially enhances the probability of co-occurrence of EPEs even for distant locations. The clustering clearly demonstrates the prevalence of distinct synoptic-scale atmospheric conditions during the occurrence of EPEs for different locations within the region. Results indicate that clustering based on a combination of SLP and Z500 can increase the conditional probability of EPEs by more than three (3) times (median value for all grid cells) from the nominal probability of 1% for the P99 EPEs. Such strong links can support extended-range forecasts.

## 1. Introduction

The Mediterranean lies at the crossroads of Africa, Asia, and Europe. It is a medium-scale coupled atmosphere-ocean system of unique character; the result of complex topography, orographic influences and interactions with large-scale water and land bodies around the domain. As the Mediterranean extends within the Hadley and Ferrel cells of the northern hemisphere, it experiences influence of both midlatitude and tropical climate variability (Trigo et al., 2006).

One of the most frequent natural hazards that affects the domain is Extreme Precipitation Events (EPEs), which can lead to landslides and floods (pluvial, fluvial, flash). Such events have severe negative consequences in society, environment and economy (Jonkman, 2005). EPEs are identified as the meteorological hazard of highest negative impact for many of the Mediterranean regions, given their frequency and the high vulnerability of the densely populated areas (Llasat et al., 2010, 2013). Moreover, the Mediterranean is a region where ongoing climate change is expected to have high impacts. Therefore, it is defined as "hot spot" to this regard (Giorgi, 2006). Some of the projected changes include increasing frequency and magnitude of EPEs (Cardell et al., 2020; Frei et al., 1998; Gao et al., 2006; Toreti et al., 2013). In fact, recent studies have already confirmed such changes using observational data (e.g. Alexander et al., 2006; Kostopoulou and Jones, 2005; Papalexiou and Montanari, 2019; Vautard et al., 2015). These changes, together with high vulnerability of the domain and high economic assets in many coastal areas, are expected to lead to some of the most dramatic increases globally in average annual losses due to flooding by 2050 (Hallegatte et al., 2013). Thus, there are large efforts and ongoing research to better understand this natural hazard and to identify ways of improving EPE predictability as well as to increase resilience of affected regions and societies (e.g. Hydrological Mediterranean Experiment – HyMeX; Drobinski et al., 2014). Such advances are of key relevance to mitigate adverse impacts and related risks.

Previous research has identified the spatiotemporal characteristics of EPEs over the Mediterranean, with most events occurring during winter half years (e.g. Grazzini et al., 2020; Houssos and Bartzokas, 2006; Khodayar et al., 2018; Lolis and Türkeş, 2016; Merino et al., 2016; Pavan et al., 2019). These results agree with the accumulated patterns over the domain, as those months record the highest precipitation amounts (Mariotti et al., 2002). In respect to EPEs, there is a clear seasonal differentiation between the western and eastern parts of the domain; with most events occurring during autumn and winter, respectively; results that are consistent with both reanalysis and observation datasets (Cavicchia et al., 2018; Raveh-Rubin and Wernli, 2015).

Many of the EPEs over the Mediterranean are closely associated with synoptic-scale atmospheric flow patterns, such as cut-off lows, throughs, and warm/cold fronts (Merino et al., 2016; Toreti et al., 2010, 2016), while others are also related to Mesoscale Convective Systems (Rigo et al., 2019). Some of the EPEs are moreover connected to cyclonic formations over the Mediterranean Sea (Lionello et al., 2006), also known as Medicanes (Cavicchia et al., 2014). The quantification of such links is crucial, as Numerical Weather Prediction (NWP) models are more skillful in predicting large-scale circulations rather than localized EPEs, especially for extended-range forecasts (Lavaysse et al., 2018; Lavers et al., 2017, 2018; Vitart, 2014). One way to statistically identify EPE connections with large-scale patterns is to analyse EPE composites for various atmospheric variables and at different vertical levels. Toreti et al. (2016) presented such connections by analysing the field of potential vorticity, while Merino et al. (2016) used a range of fields describing the dynamics of the atmosphere in the low- and mid-level troposphere, an approach also used by Toreti et al. (2010).

Recently, progress has also been achieved in connecting EPEs and general precipitation regimes over the Mediterranean with climatic modes of variability, either over the specific domain, or over extended regions covering for example the Euro-Atlantic region. Detailed reviews about such connections is presented in Alpert et al. (2006), Trigo et al. (2006) and Xoplaki et al. (2012). Vicente-Serrano et al. (2009) analysed relationships between occurrence and magnitude of EPEs over northeast Spain and North Atlantic Oscillation (NAO; Walker and Bliss, 1932), Western Mediterranean Oscillation (WeMO; Martin-Vide and Lopez-Bustins, 2006), and Mediterranean Oscillation (MO; Conte et al., 1989). They found that the most extreme daily precipitation during winter is expected for negative WeMO events. The connections are stronger when analysing the impact of such oscillations over aggregated data. NAO, for example, is found to be significantly associated with winter precipitation not only for the west Mediterranean, but also for its eastern part (Quadrelli et al., 2001; Türkeş and Erlat, 2005).

Given the existing research within the area, the aim of this work is two-fold. This study focuses a) on better understanding spatiotemporal variability of EPEs over the Mediterranean, using ERA5, the latest reanalysis data from ECMWF. Data extending up to the recent period allow to update the findings and to compare the results with similar studies. Moreover, physical consistency, fine spatiotemporal resolution, and completeness of the dataset (compared to rain-gauge observations that have frequently many missing and/or unreliable data, or inconsistencies among different countries/regions) allow to analyse EPEs at fine spatial scales. This work aims b) at quantifying the connections between EPEs and large-scale atmospheric flow patterns over the Mediterranean domain. The existing literature mainly investigates such connections by analysing EPE composites derived from in-situ measurements, leading to reduced spatial coverage and use of limited temporal information. Thus, by using the entire daily atmospheric variability and the ERA5 dataset, a holistic overview of such connections over the whole domain can be delivered. Finally, this study aims at using the derived information in future research on sub-seasonal predictability of the identified patterns.

The data and methods used for this study, are described in sections 2 and 3, respectively. Section 4 presents the results and discusses the findings, and finally, section 5 summarizes the main conclusions and points out limitations and suggestions for future pathways.

## 2. Data

Data from ERA5, the latest reanalysis dataset of ECMWF (Hersbach et al., 2020), are used for the years 1979–2019, which provide a complete record. The data are generated at hourly resolution in a horizontal grid of roughly 30 km x 30 km, using vertical 137 levels from the surface up to 80 km height to resolve the atmosphere. Four different variables of the dataset are used in this study: Total precipitation, sea level pressure (SLP), temperature at 850 hPa (T850), and geopotential height at 500 hPa (Z500).

Total precipitation is analyzed at a horizontal resolution of $0.25°$ x $0.25°$, to derive statistical information of EPEs at fine spatial scales. This resolution is moreover the closest to the original one. The selected domain covers the area $29°/47°N$ and $-8°/38°E$, referred to as **Mediterranean** from now on. Precipitation data within ERA5 are calculated from short-term forecasts of hourly accumulations. To reduce the influence of possible spin-up errors of the forecast model outputs (Dee et al., 2011), daily precipitation is calculated based on the accumulation of the forecast steps 7–18 for the models initiated at 18:00 UTC of the previous day, and at 06:00 UTC of the target day. This method results in a spin-up time of 6 hours before incorporating any of the model outputs into the analysis.

In order to understand the connection of large-scale atmospheric patterns to the occurrence of EPEs, atmospheric variability in the lower and middle troposphere is analyzed, based on SLP, T850 and Z500. These variables have been selected, as previous work demonstrated their importance in defining synoptic environments and fronts that are connected to EPEs (e.g. Catto et al., 2014; Greco et al.,

2020; Hidalgo-Muñoz et al., 2011; Xoplaki et al., 2004). The selected spatial resolution for these variables is 1º x 1º. Mean daily values were calculated by averaging all 24 available hourly data, and the spatial domain used for deriving the data is 26º/50ºN and -11º/41ºE, so that the analysis captures the influence of the adjacent areas, e.g., the Atlantic Ocean and the Alps. It should be stated that during the exploratory analysis, various domains of larger spatial extend were analyzed (e.g. Euro Atlantic domain, Mediterranean further extended to the Atlantic), with connections between EPEs and large-scale patterns being weaker. This corroborates previous studies, demonstrating that larger domains are not efficient to optimize relationship between circulation types and precipitation (Beck et al., 2016).

## 3. Methodology

### a. Extreme Precipitation Events (EPEs) and spatiotemporal analysis

EPEs are identified based on the 99$^{th}$ percentile (P99) of the daily distribution, considering all available data for each grid cell. It should be stated that EPEs defined on seasonal data percentiles are not considered, as the aim of this work is to analyse EPEs that are closely associated with negative consequences to society. Moreover, because of significant differences in precipitation amounts and number of dry days between the various locations of the domain (especially north vs south and highlands vs lowlands), wet-days-derived EPEs and fixed-threshold-derived EPEs are not given preference. It is worth noting that during the conducted analysis percentiles of lower frequency (e.g. P95) were also tested, providing no substantial differences in the conclusions. Thus, the presented results are solely based on the P99 EPEs analysis.

The identified events are analyzed based on their seasonality, in order to quantify the importance of seasons in the occurrence of EPEs. As the data used cover the period January 1979 – December 2019, the available number of months is the same for all seasons (4 x 3 months x 41 years).

The degree of temporal dependence is a crucial factor, as multiple EPEs within short temporal intervals can significantly increase flood and landslide risks. In this work, four different temporal intervals are analyzed. The EPE percentage is calculated at each grid cell that occurred within the selected interval from the preceding EPE at the same grid cell. The intervals are 1, 3, 7, and 15 days, to understand the persistence of EPEs at short, medium and extended range temporal scales. Each of these scales yields different impacts and is associated with different meteorological and climatological drivers. Other intervals have been tested and would not lead to different conclusions.

### b. Large-scale patterns

EOF analysis on the covariance matrix (Wilks, 2011) is performed on the derived daily anomalies, independently for each of the selected variables. These anomalies are from now on simply referred to by the variable name. The square root of the cosine of latitude is used for weighting the data and giving equal-area weighting on the covariance matrix. The climatology of these variables is calculated with a 5-day smoothing window. To derive the anomalies, each day of the 1979–2019 period was subtracted from its corresponding daily climatology. The necessary number of modes (principal components) that explain at least 90% of the total variance, was retained to obtain a compressed dataset that provides most of the available information.

Non-hierarchical K-means clustering (Hartigan and Wong, 1979) is implemented on the data projections on the retained principal components, and the Euclidean distance is used as the similarity measure. The analysis is conducted with the Python package *scikit-learn* (Pedregosa et al., 2011). Clustering is performed independently for each variable, as well as for all different combinations of the three variables used. For the latter, the projections are pre-processed to show comparable units, and weights relevant to the explained variance. More specifically, standardization was implemented so that all projections have a standard deviation of 1. In the next step, each projection was multiplied with the square root of the percentage of the total variance of the corresponding variable that it explains, so that the projections are weighted based on their relative importance. The 95% uncertainty intervals are computed according to Lee et al. (2019), using cluster persistence and the effective sample size (Wilks, 2011) based on self-transition probability.

The number of clusters can be constrained and selected based on (semi)objective criteria, but there is also a level of subjectivity introduced (Gong and Richman, 1995). Moreover, the particular use of the derived clusters is crucial for the final selection. Given that this work aims at conditioning EPEs for further use of the results on subseasonal forecasting, a large number of clusters is not recommended, since only a broad indication of the expected synoptic-scale pattern is possible for such timescales (Neal et al., 2016). Therefore, K-means clustering is performed by generating 7 up to 12 clusters. Smaller number of clusters are not as useful, since connections to EPEs (explained below) are weaker.

### c. Connection of large-scale patterns to EPEs

The importance of each cluster in EPE occurrence was assessed, based on the conditional probability of EPEs for each cluster. In this context, we refer to the largest conditional probability over all clusters (for each combination of used variables and selected number of clusters) as the Maximum Conditional Probability (MCP). The conditional probability is highly beneficial for this study since it considers any potential differences in the occurrence probability of the identified regimes. Moreover, it can be directly compared with the nominal probability of the selected EPEs (meaning 1%). Furthermore, to obtain a more holistic understanding of the classification benefits for the different variables and the number of clusters, additional indicators were analyzed, namely the EPE percentage allocated to each cluster (e.g. as in Yiou and Nogaj, 2004), and the percentage of grid cells that exhibit statistically significant connections with at least one cluster. Statistically significant connections between EPEs and clusters are assessed with the two-tailed 95% confidence interval of binomial distribution (e.g. Olmo et al., 2020). Although the interest of this work lies on positive relationships between EPEs and clusters, two-tailed test is preferred over one-tailed test, as the latter has higher type 1 errors. The probability of occurrence that is introduced for the significance testing is the upper $95^{th}$ confidence interval, so that strict criteria are used due to the inherent uncertainties associated with clustering.

## 4. Results and Discussion

### a. Spatiotemporal characteristics

As expected, results indicate the strong influence of orography (Figure 1a) on precipitation intensity (e.g. Atlas Mountains, Alps, coast of west Balkans), as well as the importance of latitude, with locations closer to/on the sinking air masses of the Hadley cell (north Africa) having significantly smaller thresholds than the locations at northern latitudes. Figure 1b presents the derived P99 EPEs thresholds within the domain. The spatial EPE pattern, as well as their magnitudes are very similar to the study of Cavicchia et al. (2018) who used the E-OBS dataset (gridded dataset from observational data) with same spatial resolution, and identified wet-days $99^{th}$ percentile intensities. The greater intensity differences in the southern Mediterranean compared to the rest of the domain can be attributed to the large number of dry days in that region. Since only the wet days were used in the work of Cavicchia et al. (2018), it is expected that the derived percentile magnitudes over those areas will be larger compared to the current study that uses all daily values.

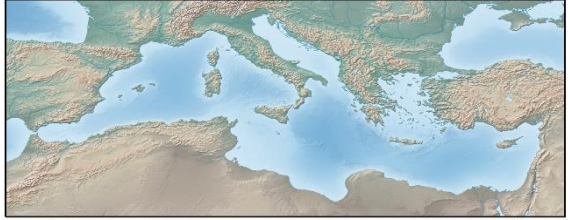
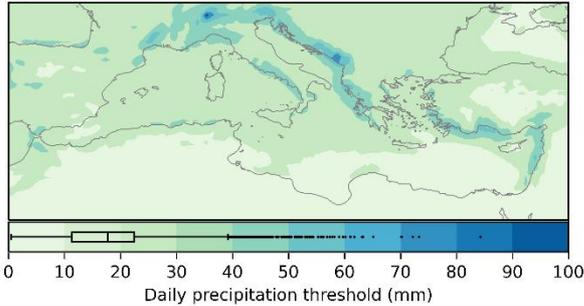

*Figure 1 Orography (a), and P99 (b) EPEs threshold for the study area. The orography subplot (a) uses multiple color scales for distinguishing elevation and climatic differences (arid vs humid). More information about the color scales can be found at www.naturalearthdata.com where the data were downloaded from (file name "HYP_LR_OB_DR.tif"). The boxplot at subplot (b) presents the P99 distribution of all grid cells, indicating the median value and extending from the lower to upper quartile. The whiskers extend to the further available value up to 1.5*IQR from the lower and upper quartile, and all other values outside this range are presented as outliers.*

The EPE seasonality demonstrates a west/east divide, with most EPEs occurring during winter (autumn) for the eastern (western) parts of the domain (Figure 2, Figure S1**Error! Reference source not found.**). These findings agree with results from other studies using different datasets (Cavicchia et al., 2018; Raveh-Rubin and Wernli, 2015), and indicate that different synoptic-scale configurations are likely to generate EPEs at the different regions of the domain (Raveh-Rubin and Wernli, 2015). More than 70% of the EPEs in parts of southeast Mediterranean occur during winter, whereas for parts of west Mediterranean, Italy and west Balkans, over 60% of events occur during autumn. These two seasons interchange between 1$^{st}$ and 2$^{nd}$ place in terms of EPE occurrence for most of the domain. The north Balkans and southeast Europe exhibit a different pattern, with summer and spring being the two seasons of the highest EPE occurrence for most of the area. This indicates that the Mediterranean Sea has less influence in those areas, where EPEs are associated with different large-scale patterns. This can also be explained by orography, with the mountain ranges over the south Balkans minimize direct interactions between the Mediterranean Sea and north Balkans/southeast Europe. Spring and summer are important seasons for mountainous locations within the domain (e.g. Alps, Pyrenees). The above can be attributed to EPEs of convective nature that occur in such areas (Romero et al., 2001). These events, resulting from thermal low pressure systems over the region (Campins et al., 2000), are further enhanced by orography. Because of the generally small spatial scale of such convective events, this was not explored in the study of Raveh-Rubin and Wernli (2015), who used ERA-Interim with 1$^\circ$ spatial resolution and additionally implemented spatiotemporal smoothing. This shows that recent advances in the resolution of reanalysis data, due to increased computational efficiency and computation power, can bring additional benefits and more realistic information even for finer scales. Finally, spring is also important for north Africa and is the season of highest/2$^{nd}$ highest occurrence of EPEs for most of this area.

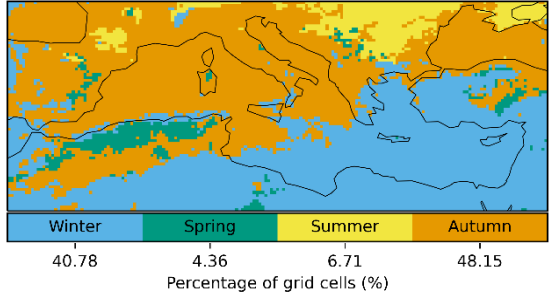

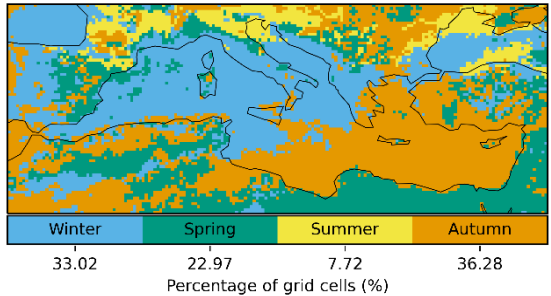

*Figure 2 Season of highest (a) and 2nd highest (b) occurrence of P99 EPEs per grid cell.*

Temporal EPE dependence is strong even for the 1-day interval, especially so for the dry parts of the domain in northeast Africa and adjacent parts of the Mediterranean Sea (Figure 3). Orography enhances such dependencies, with locations in Atlas Mountains, Alps, and west Balkans coast, differentiating from their neighboring locations in many of the presented results. This can be attributed to orographic lifting and forced convection that occurs on the windward side of the mountains. These processes can trigger EPEs even when large-scale systems are in distant areas, as long as moisture advection is directed towards the mountains (Pfahl, 2014). Thus, for 1- and 3-day intervals, which are associated with eastward propagation of synoptic-scale weather, these regions strongly differ from surrounding locations. The event percentage median values that occur within 1 day of a preceding P99 EPE is about 11%. These values increase to about 21% for 7 days interval, and almost 30% for 15 days interval. Such strong dependencies indicate persistent meteorological (e.g. troughs, cut-off lows, storm tracks, cyclones) and climatological conditions (e.g. weather regimes; Vautard, 1990, Barnes and Hartmann, 2010).

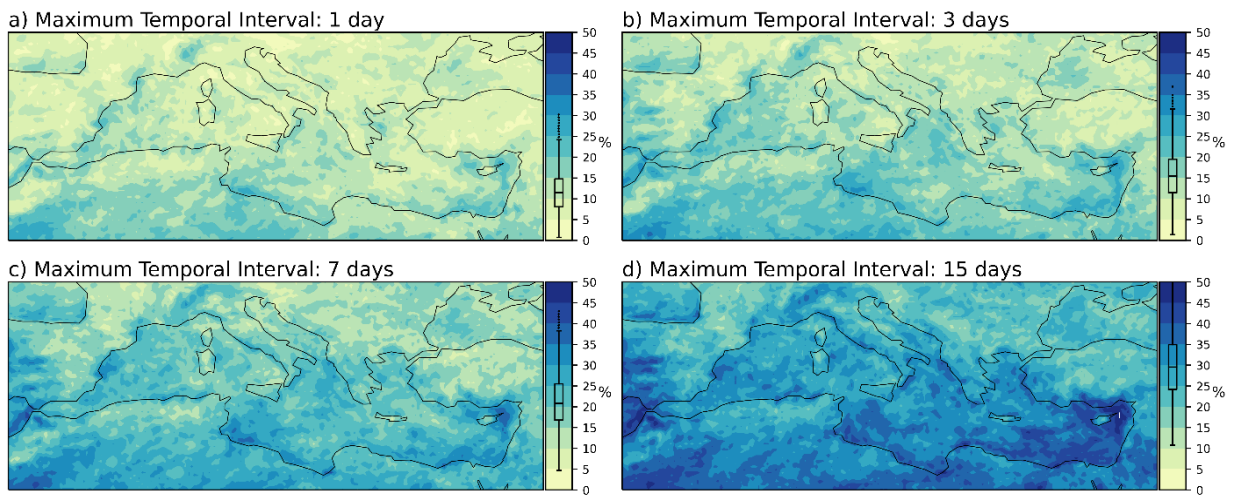

*Figure 3 Temporal Dependencies of EPEs. The figures present the percentage of P99 EPEs that occur within the selected temporal intervals (1 day: a; 3 days: b; 7 days: c; 15 days: d) from a preceding EPE at the same grid cell. The boxplots at the subplots are as in Figure 1.*

b. Connection to large-scale patterns

Before presenting EPE connections to large-scale patterns, it is worth commenting on the EOFs for Z500, with the first 4 of them explaining more than 77% of total variance (Figure S2). The first component explains one third of the total variance. It shows that east and west Mediterranean have opposite behaviour, with the core of the EOF being located over France. This east/west divide that is additionally depicted in the EPEs analysis, is also noticeable in the 2$^{nd}$ EOF, while the 3$^{rd}$ EOF exhibits an omega pattern with the central parts of the domain having an opposite behaviour from the east and west subdomains. These 4 EOFs are associated with the Mediterranean Oscillation (MO; difference in SLP anomalies between Gibraltar and Lod, Israel) and WeMO (difference in SLP anomalies between San Fernando, Spain and Padova, Italy). It should be stated that EOF analysis based on winter half and summer half year, had no substantial differences in the patterns and the percentage of expected variance, while the EOF based on SLP has similar patterns, but is influenced by the orography (not shown).

Figure 4 presents the results on the connection of P99 EPEs to the identified weather regimes (clusters) for all studied variables and number of clusters. Figure 4a presents the Maximum Conditional Probability (MCP) and Figure 4b refers to the EPE percentages that occur in the cluster of MCP for each grid cell. Both plots show the median value from all grid cells with statistically significant connections to the generated weather regimes. As can be noticed from Figure 4a, the conditional probability is increased for all variables by increasing the number of clusters. Clustering based on the combined information from SLP and Z500 outperforms all other clustering results. The other variables perform similarly, except for clustering based on T850, which is substantially worse. This is expected, as many EPEs over the Mediterranean are connected to troughs and cut-off lows, which mainly have a strong signal in geopotential height and surface pressure. Such patterns also show a signal in the temperature fields due to the frequent generation of cold and warm fronts. Yet, as these formations are of smaller spatial extent, clustering based solely on temperature fields for such a large domain is less effective compared to clustering based on the other variables.

The EPE percentage associated with the cluster of MCP decreases with a higher number of clusters. Besides SLP and T850, which have a weak connection, all other variables perform relatively similarly. This tradeoff between conditional probability and percentage of EPEs is expected, as by increasing the number of clusters, the connection between EPEs and some of the derived clusters becomes stronger. Yet, at the same time, as clusters correspond to smaller number of days, the associated EPE percentages decrease. It can be noticed that for SLP-Z500 combination the MCP saturates from 9 to 12 clusters. At the same time, 9 clusters perform very similarly to 8 in terms of EPE percentages associated with the MCP cluster. Thus, the combination of SLP and Z500 for 9-clusters K-means clustering is selected as the preferred classifier to connect EPEs to large-scale patterns. This selection is further justified by results shown in Figure S3, presenting two additional indicators for the clusters and variables studied. These are the percentage of grid cells that have statistically significant connections with at least 1 cluster (subplot a), and the percentage of EPEs (median value from all grid cells with statistically significant connections) that are significantly connected with any of the clusters (subplot b). For both indicators, the order of magnitude of all variables is generally the same and results do not change substantially by changing the number of clusters. This means that the final selection of the preferred combination is not affected by these two indicators. It is worth noting that at least half of the total P99 EPEs are significantly associated with preferential weather regimes for most of the grid cells and for all number of clusters and selected variables (expect SLP and T850). Finally, for the selected combination of SLP-Z500 and 9 clusters, the generated composites (cluster centroids, defined by averaging all data corresponding to each cluster) do not change substantially in spatial pattern and magnitude when more clusters are considered (not shown). This indicates that the clusters demonstrate coherent similarities between individual samples. Detailed results about this classifier are presented below.

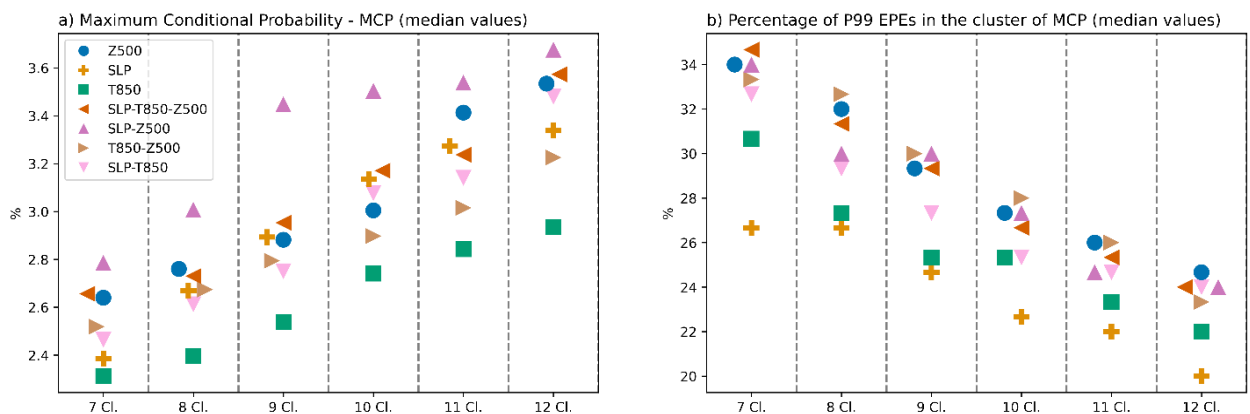

*Figure 4 a) Maximum Conditional Probability (MCP) of P99 EPEs for the studied variables and number of clusters. b) Percentage of P99 EPEs associated with the cluster corresponding to MCP. For both plots, values represent the median value of all grid cells that have statistically significant connection to the generated weather regimes.*

Figure 5 presents the composites of the 9-classes K-means clustering based on the combined SLP (colour shading) and Z500 (contours) fields. The derived composites have a naming convention based on the location of negative anomalies (if existing), so that the connections between clusters and EPEs are more intuitive. The composites exhibit noticeable differences in type, magnitude, or location of the large-scale patterns. The importance of the Atlantic, and the storms generated over that area is very clear. It can be noticed that clusters 1 (Atlantic Low), 2 (Biscay Low), and 3 (Iberian Low) are associated with negative anomalies that relate to unstable conditions centred over/near the Atlantic. Cluster 4 (Sicilian Low) has a negative anomaly centred over the central Mediterranean, although of low magnitude, and for cluster 5 (Balkan Low) the negative anomalies are centred over (west) Balkans. Cluster 6 (Black Sea Low) has a dipole structure with positive anomalies over west Mediterranean and negative over east, while cluster 7 (Mediterranean High) corresponds to positive anomalies and stable conditions over the whole domain. Finally, clusters 8 (Minor Low) and 9 (Minor High) correspond to negative and positive anomalies of low magnitude over most of the domain and are associated with days that do not indicate distinct cyclonic or anticyclonic conditions of synoptic scale over the area. They can be therefore considered as the "no-anomaly" clusters in the studied domain.

Seasonal and annual variability in the occurrence of most clusters is very high (Figure S4). Atlantic Low and Sicilian Low have median occurrences of similar magnitude throughout the different seasons, while Minor Low, Minor High, and Mediterranean High show very high seasonal variability. It should be noted that Mediterranean High is very frequent during winter. This is crucial for the occurrence of cold spells (not studied in this work), which can be generated when such conditions persists for many days (Ferranti et al., 2018; Grams et al., 2017). Biscay Low, Iberian Low, Balkan Low, Black Sea Low and Mediterranean High, with negative/positive anomalies over large parts of the domain, barely occur in summer. Atmospheric characteristics associated with these clusters are mainly driven by storm tracks and unstable conditions that prevail during winter and the intermediate seasons of autumn and spring. The high annual variability for all clusters indicates that their occurrence is modulated by climatic variability and larger-scale phenomena, e.g. NAO (e.g. Dünkeloh and Jacobeit, 2003; Xoplaki et al., 2012).

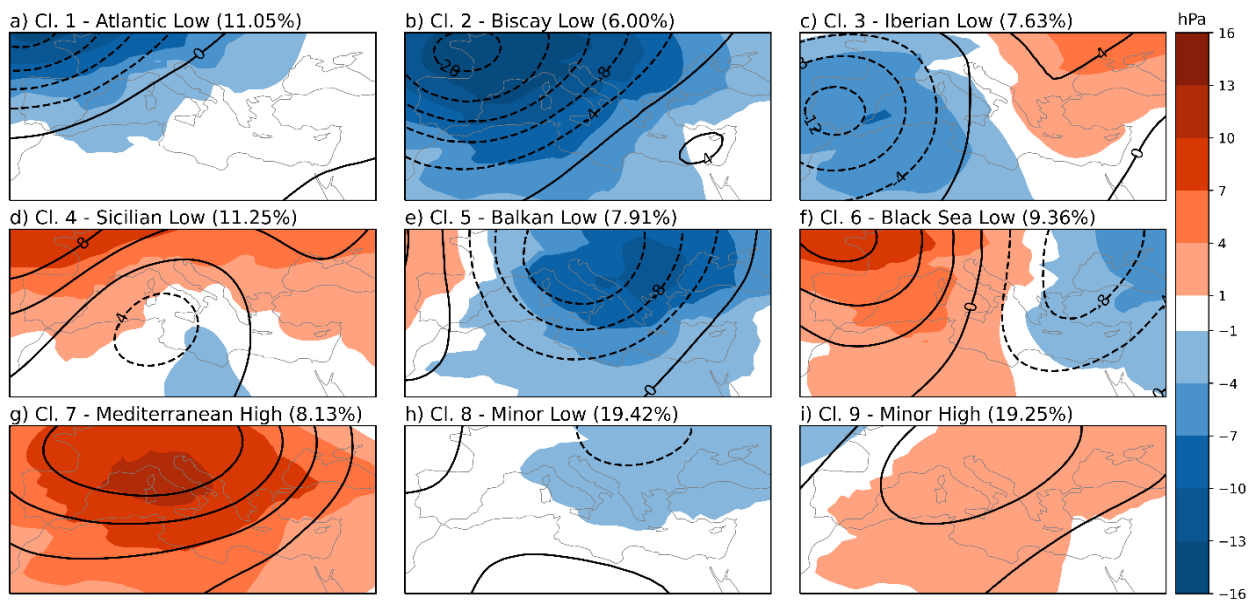

*Figure 5 Composite of clusters, derived with K-means clustering on the principal components' projections of SLP and Z500 anomalies. Color shading refers to SLP anomalies (hPa), and contours to Z500 anomalies (dm). Percentages indicate the occurrence of each cluster to the total number of available days.*

Statistically significant clusters of highest (MCP; a) and 2$^{nd}$ highest (b) conditional probability of P99 EPEs for each grid cell are presented in Figure 6. The top panels present the associated cluster. The legend indicates the percentage of grid cells allocated to each cluster. Subplots c and d present the corresponding conditional probabilities, and subplots e and f the percentage of total EPEs for each grid cell that occur during the associated cluster.

The clusters corresponding to the highest conditional probability for each grid cell show very smooth spatial behaviour, with neighbouring regions allocated mainly to the same cluster. Few discontinuities (e.g. west Italy and west Balkans vs East Italy, west Greece and west Turkey vs east Greece and Aegean) are mainly associated with orography (explained below). Only 0.38% of the total grid cells (located at the edge of the domain in the Middle East) cannot be significantly related with any of the derived weather regimes, while this percentage increases to 22.96% for the second "best" cluster. Results for the 2$^{nd}$ highest conditional probability are patchier, yet closely relate to the results of subplot a. Less than 25% of the total grid cells are significantly associated with more than 2 clusters (not shown), with conditional probabilities for each weather regime at the statistically significant grid cells being presented in Figure S5.

Cluster composites can physically explain connections with EPEs. Negative anomalies of pressure and geopotential height centred over the Atlantic (Atlantic Low) correspond to high chance of EPEs over the west Iberian Peninsula and southern France. As these anomalies move further east (Biscay Low and Iberian Low), the most impacted locations are the west and west-central Mediterranean. Iberian Low has significant connections over a large area. It is the cluster of the highest/2$^{nd}$ highest conditional probabilities for regions spanning from the Atlantic (west Mediterranean) to the west Balkans (east Mediterranean), and for both high and low latitudes. Moreover, the significant importance of orography is demonstrated in this cluster. As cyclonic flow and existing moisture approaches the mountains of Picos de Europa, Atlas, Alps, Apennines, and west Balkans, the windward locations are preferentially affected by EPEs, and the conditional probability for EPEs over such regions is more than 7 times higher from the nominal probability of 1%. Such results can be further explained by the importance of the moisture advection towards the mountains in the generation of extreme precipitation (Pfahl, 2014). Such orographic influence is also demonstrated in the association of Sicilian Low with EPEs. In this cluster, the central Mediterranean, and locations east of the Apennines and east of the mountain ranges of the west Balkans (Dinaric Alps, Pindos) are affected. Balkan Low and Black Sea Low are highly associated with EPEs over the Balkans and Turkey, respectively, while Mediterranean High, with the anticyclonic flow centred over the domain, generates EPEs over eastern north Africa and the Middle East. Minor Low and Minor High show minor association with EPEs over the domain.

The conditional probabilities associated with the MCP cluster are almost 50% higher compared to the 2$^{nd}$ cluster in general, with the latter still showing twice the probabilities compared to the nominal of 1% for most grid cells. Moreover, it can be noticed from subplots e and f that for most grid cells 30% (20%) of the total EPEs occur during the cluster of MCP (2$^{nd}$ MCP). This indicates that such classification does not only significantly increase the conditional probabilities but can also explain a large amount of the total EPEs, even when using only 2 of the 9 clusters.

Coming back to the results of clusters of highest conditional probability (Figure 6a), it is worth noting the connection between western Italy and the west Balkan coast. Not only do these regions correspond to the same cluster, but there is also a high degree of temporal overlap (Figure 7). More than 30% of EPEs over parts of west Italy occur on the same day as EPEs over the coast of Croatia and Montenegro (and nearby areas). This is the effect of the Apennines that block the westerly airflow and force the moisture to precipitate in west Italy. Thus, a strong connection in the co-occurrence of EPEs between west Italy and west Balkans is established, whereas east Italy and the Adriatic Sea have a clear dissociation. It can be concluded that orography is a key modulator of the spatiotemporal connections between locations.

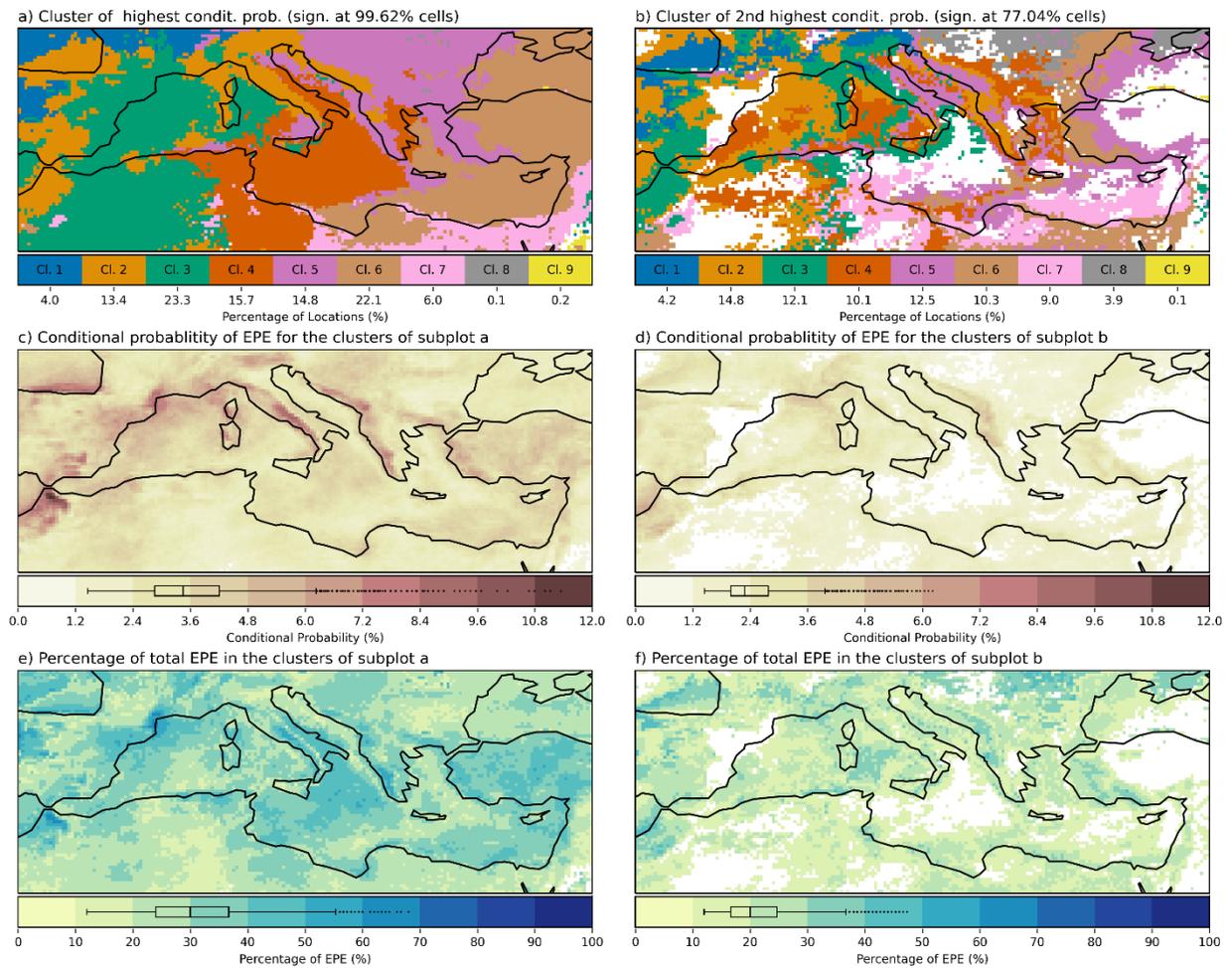

*Figure 6 Connection of P99 EPEs and weather regimes: a) and b) depict the clusters of highest and 2nd highest occurrence probability for each grid cell respectively; c) , d) associated conditional probability of the clusters at subplots a) and b), respectively; e), f) associated percentage of total EPEs occurring at the clusters of subplots a) and b) respectively. The boxplots at subplots c)-f) are as in figure 1, but only for statistically significant grid cells.*

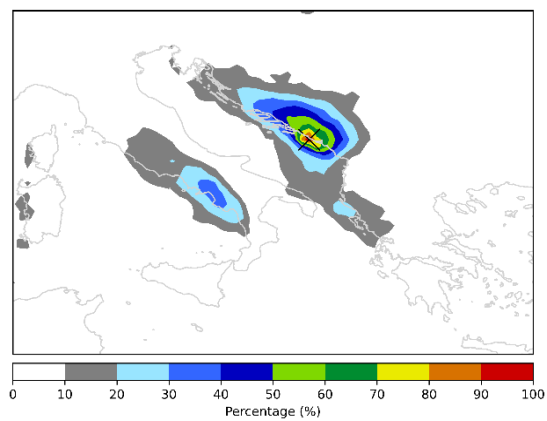

*Figure 7 Cooccurrence of P99 EPEs between the coast of Croatia/Montenegro (black cross) and the rest of the locations*

## 5. Conclusions

Spatiotemporal characteristics of daily EPEs (being defined based on the 99$^{th}$ percentile, P99) over the Mediterranean are analyzed, using the ERA5 dataset. Their connection to large-scale atmospheric flow patterns is quantified by using EOF analysis and subsequent K-means clustering of the atmospheric variability in the lower and middle troposphere (SLP, T850, and Z500 anomalies).

The results indicate that the Mediterranean region is divided into two domains in terms of EPE seasonality, with autumn being the prevailing season for the western domain, and winter for the eastern one. Spatiotemporal connections between EPEs are very strong and are furthermore modulated by orography and regional climate. For most of the analyzed grid cells, at least 20% of their EPEs occur up to 7 days after a preceding EPE at the same grid cell. This indicates persistent meteorological conditions. Such connections are even stronger at high-altitude locations (e.g., Alps and Atlas Mountains), as well as regions characterized by their dry climatic conditions (north Africa and Middle East). West Italy and the west Balkans demonstrate a remarkable temporal EPE connection with more than 20% of EPEs occurring on the same day; an effect of the Apennine mountains.

Clustering, based on selected atmospheric variables, demonstrates that a combination of SLP and Z500 anomalies leads to the highest association of EPEs with the derived clusters. 9-class K-means clustering is the preferred classifier. Clusters correspond to negative or positive anomalies over west, central, or east Mediterranean, and can be associated with cyclonic and anticyclonic conditions on the synoptic scale. These clusters show a clear connection with the observed EPEs over the vast majority of the studied domain; a relationship that can be explained by atmospheric dynamics associated with each cluster. The conditional EPE probability increases by more than 3 times for most locations, compared to the nominal probability of 1% for the studied EPEs. Additionally, more than 30% of the P99 EPEs preferentially occur during the cluster of highest conditional probability for most grid cells. Orography further enhances these relationships. Locations in windward parts of the Apennine mountains, Picos de Europa, and Atlas Mountains, show MCP of P99 EPEs over 7%, and more than 40% of the EPEs occur during the associated cluster.

Information from this study delivers additional benefits when analysing EPE predictability at extended-range forecasts. It has been demonstrated that many EPEs can be associated to large-scale atmospheric flow patterns. This is important, since NWP models are more skillful in predicting such large-scale patterns over extended-range forecasts. Optimal selection of weather clustering does also depend on NWP model performance for the selected lead time. Various combinations of domain extent, number of clusters, and atmospheric variables should be further analyzed to maximize the usefulness of the currently-used NWP models, when assessing the performance of extended-range forecasts. This will be studied in future works. This work uses ERA5 reanalysis data to delineate EPEs. ERA5 is produced using a high-resolution model that incorporates state of the art physics and assimilates multiple observations to calculate the various atmospheric variables. Despite challenges of reanalysis data (e.g. limitations mentioned at Hersbach et al., 2020), this provides the opportunity to use a physically-consistent dataset for the analysis performed. As precipitation is a complex phenomenon, highly varying across spatiotemporal scales, it would be useful to perform this analysis using observational data and assess whether finer spatiotemporal resolutions have similar characteristics.

## Acknowledgements

This work is part of the EU International Training Network (ITN) Climate Advanced Forecasting of sub-seasonal Extremes (CAFE). The project receives funding from the European Union's Horizon 2020 research and innovation programme under the Marie Skłodowska-Curie Grant Agreement No 813844. The authors would like to thank David Richardson for his fruitful suggestions for improving the quality of this manuscript.

## References

Alexander, L.V., Zhang, X., Peterson, T.C., Caesar, J., Gleason, B., Tank, A.M.G.K., Haylock, M., Collins, D., Trewin, B., Rahimzadeh, F., Tagipour, A., Kumar, K.R., Revadekar, J., Griffiths, G., Vincent, L., Stephenson, D.B., Burn, J., Aguilar, E., Brunet, M., Taylor, M., New, M., Zhai, P., Rusticucci, M., Vazquez-Aguirre, J.L., 2006. Global observed changes in daily climate extremes of temperature and precipitation. J. Geophys. Res. Atmospheres 111. https://doi.org/10.1029/2005JD006290

Alpert, P., Baldi, M., Ilani, R., Krichak, S., Price, C., Rodó, X., Saaroni, H., Ziv, B., Kishcha, P., Barkan, J., Mariotti, A., Xoplaki, E., 2006. Chapter 2 Relations between climate variability in the Mediterranean region and the tropics: ENSO, South Asian and African monsoons, hurricanes and Saharan dust, in: Lionello, P., Malanotte-Rizzoli, P., Boscolo, R. (Eds.), Developments in Earth and Environmental Sciences, Mediterranean. Elsevier, pp. 149–177. https://doi.org/10.1016/S1571-9197(06)80005-4

Barnes, E.A., Hartmann, D.L., 2010. Dynamical Feedbacks and the Persistence of the NAO. J. Atmospheric Sci. 67, 851–865. https://doi.org/10.1175/2009JAS3193.1

Beck, C., Philipp, A., Streicher, F., 2016. The effect of domain size on the relationship between circulation type classifications and surface climate. Int. J. Climatol. 36, 2692–2709. https://doi.org/10.1002/joc.3688

Campins, J., Genovés, A., Jansà, A., Guijarro, J.A., Ramis, C., 2000. A catalogue and a classification of surface cyclones for the Western Mediterranean. Int. J. Climatol. 20, 969–984. https://doi.org/10.1002/1097-0088(200007)20:9<969::AID-JOC519>3.0.CO;2-4

Cardell, M.F., Amengual, A., Romero, R., Ramis, C., 2020. Future extremes of temperature and precipitation in Europe derived from a combination of dynamical and statistical approaches. Int. J. Climatol. 1–28. https://doi.org/10.1002/joc.6490

Catto, J.L., Nicholls, N., Jakob, C., Shelton, K.L., 2014. Atmospheric fronts in current and future climates. Geophys. Res. Lett. 41, 7642–7650. https://doi.org/10.1002/2014GL061943

Cavicchia, L., Scoccimarro, E., Gualdi, S., Marson, P., Ahrens, B., Berthou, S., Conte, D., Dell'Aquila, A., Drobinski, P., Djurdjevic, V., Dubois, C., Gallardo, C., Li, L., Oddo, P., Sanna, A., Torma, C., 2018. Mediterranean extreme precipitation: a multi-model assessment. Clim. Dyn. 51, 901–913. https://doi.org/10.1007/s00382-016-3245-x

Cavicchia, L., von Storch, H., Gualdi, S., 2014. A long-term climatology of medicanes. Clim. Dyn. 43, 1183–1195. https://doi.org/10.1007/s00382-013-1893-7

Conte, M., Giuffrida, A., Tedesco, S., 1989. The mediterranean oscillation : impact on precipitation and hydrology in Italy 121–137.

Dee, D.P., Uppala, S.M., Simmons, A.J., Berrisford, P., Poli, P., Kobayashi, S., Andrae, U., Balmaseda, M.A., Balsamo, G., Bauer, P., Bechtold, P., Beljaars, A.C.M., Berg, L. van de, Bidlot, J., Bormann, N., Delsol, C., Dragani, R., Fuentes, M., Geer, A.J., Haimberger, L., Healy, S.B., Hersbach, H., Hólm, E.V., Isaksen, L., Kållberg, P., Köhler, M., Matricardi, M., McNally, A.P., Monge-Sanz, B.M., Morcrette, J.-J., Park, B.-K., Peubey, C., Rosnay, P. de, Tavolato, C., Thépaut, J.-N., Vitart, F., 2011. The ERA-Interim reanalysis: configuration and performance of the data assimilation system. Q. J. R. Meteorol. Soc. 137, 553–597. https://doi.org/10.1002/qj.828

Drobinski, P., Ducrocq, V., Alpert, P., Anagnostou, E., Béranger, K., Borga, M., Braud, I., Chanzy, A., Davolio, S., Delrieu, G., Estournel, C., Boubrahmi, N.F., Font, J., Grubišić, V., Gualdi, S., Homar,

V., Ivančan-Picek, B., Kottmeier, C., Kotroni, V., Lagouvardos, K., Lionello, P., Llasat, M.C., Ludwig, W., Lutoff, C., Mariotti, A., Richard, E., Romero, R., Rotunno, R., Roussot, O., Ruin, I., Somot, S., Taupier-Letage, I., Tintore, J., Uijlenhoet, R., Wernli, H., 2014. HyMeX: A 10-Year Multidisciplinary Program on the Mediterranean Water Cycle. Bull. Am. Meteorol. Soc. 95, 1063–1082. https://doi.org/10.1175/BAMS-D-12-00242.1

Dünkeloh, A., Jacobeit, J., 2003. Circulation dynamics of Mediterranean precipitation variability 1948–98. Int. J. Climatol. 23, 1843–1866. https://doi.org/10.1002/joc.973

Ferranti, L., Magnusson, L., Vitart, F., Richardson, D.S., 2018. How far in advance can we predict changes in large-scale flow leading to severe cold conditions over Europe? Q. J. R. Meteorol. Soc. 144, 1788–1802. https://doi.org/10.1002/qj.3341

Frei, C., Schär, C., Lüthi, D., Davies, H.C., 1998. Heavy precipitation processes in a warmer climate. Geophys. Res. Lett. 25, 1431–1434. https://doi.org/10.1029/98GL51099

Gao, X., Pal, J.S., Giorgi, F., 2006. Projected changes in mean and extreme precipitation over the Mediterranean region from a high resolution double nested RCM simulation. Geophys. Res. Lett. 33. https://doi.org/10.1029/2005GL024954

Giorgi, F., 2006. Climate change hot-spots. Geophys. Res. Lett. 33. https://doi.org/10.1029/2006GL025734

Gong, X., Richman, M.B., 1995. On the Application of Cluster Analysis to Growing Season Precipitation Data in North America East of the Rockies. J. Clim. 8, 897–931. https://doi.org/10.1175/1520-0442(1995)008<0897:OTAOCA>2.0.CO;2

Grams, C.M., Beerli, R., Pfenninger, S., Staffell, I., Wernli, H., 2017. Balancing Europe's wind-power output through spatial deployment informed by weather regimes. Nat. Clim. Change 7, 557–562. https://doi.org/10.1038/nclimate3338

Grazzini, F., Craig, G.C., Keil, C., Antolini, G., Pavan, V., 2020. Extreme precipitation events over northern Italy. Part I: A systematic classification with machine-learning techniques. Q. J. R. Meteorol. Soc. 146, 69–85. https://doi.org/10.1002/qj.3635

Greco, A., De Luca, D.L., Avolio, E., 2020. Heavy Precipitation Systems in Calabria Region (Southern Italy): High-Resolution Observed Rainfall and Large-Scale Atmospheric Pattern Analysis. Water 12, 1468. https://doi.org/10.3390/w12051468

Hallegatte, S., Green, C., Nicholls, R.J., Corfee-Morlot, J., 2013. Future flood losses in major coastal cities. Nat. Clim. Change 3, 802–806. https://doi.org/10.1038/nclimate1979

Hartigan, J.A., Wong, M.A., 1979. A K-Means Clustering Algorithm. J. R. Stat. Soc. Ser. C Appl. Stat. 28, 100–108. https://doi.org/10.2307/2346830

Hersbach, H., Bell, B., Berrisford, P., Hirahara, S., Horányi, A., Muñoz-Sabater, J., Nicolas, J., Peubey, C., Radu, R., Schepers, D., Simmons, A., Soci, C., Abdalla, S., Abellan, X., Balsamo, G., Bechtold, P., Biavati, G., Bidlot, J., Bonavita, M., Chiara, G.D., Dahlgren, P., Dee, D., Diamantakis, M., Dragani, R., Flemming, J., Forbes, R., Fuentes, M., Geer, A., Haimberger, L., Healy, S., Hogan, R.J., Hólm, E., Janisková, M., Keeley, S., Laloyaux, P., Lopez, P., Lupu, C., Radnoti, G., Rosnay, P. de, Rozum, I., Vamborg, F., Villaume, S., Thépaut, J.-N., 2020. The ERA5 global reanalysis. Q. J. R. Meteorol. Soc. n/a. https://doi.org/10.1002/qj.3803

Hidalgo-Muñoz, J.M., Argüeso, D., Gámiz-Fortis, S.R., Esteban-Parra, M.J., Castro-Díez, Y., 2011. Trends of extreme precipitation and associated synoptic patterns over the southern Iberian Peninsula. J. Hydrol. 409, 497–511. https://doi.org/10.1016/j.jhydrol.2011.08.049

Houssos, E.E., Bartzokas, A., 2006. Extreme precipitation events in NW Greece, in: Advances in Geosciences. Presented at the 7th Plinius Conference on Mediterranean Storms (2005) - 7th Plinius Conference on Mediterranean Storms, Crete, Greece, 5–7 October 2005, Copernicus GmbH, pp. 91–96. https://doi.org/10.5194/adgeo-7-91-2006

Jonkman, S.N., 2005. Global Perspectives on Loss of Human Life Caused by Floods. Nat. Hazards 34, 151–175. https://doi.org/10.1007/s11069-004-8891-3

Khodayar, S., Kalthoff, N., Kottmeier, C., 2018. Atmospheric conditions associated with heavy precipitation events in comparison to seasonal means in the western mediterranean region. Clim. Dyn. 51, 951–967. https://doi.org/10.1007/s00382-016-3058-y

Kostopoulou, E., Jones, P.D., 2005. Assessment of climate extremes in the Eastern Mediterranean. Meteorol. Atmospheric Phys. 89, 69–85. https://doi.org/10.1007/s00703-005-0122-2

Lavaysse, C., Vogt, J., Toreti, A., Carrera, M.L., Pappenberger, F., 2018. On the use of weather regimes to forecast meteorological drought over Europe. Nat. Hazards Earth Syst. Sci. 18, 3297–3309. https://doi.org/10.5194/nhess-18-3297-2018

Lavers, D.A., Richardson, D.S., Ramos, A.M., Zsoter, E., Pappenberger, F., Trigo, R.M., 2018. Earlier awareness of extreme winter precipitation across the western Iberian Peninsula. Meteorol. Appl. 25, 622–628. https://doi.org/10.1002/met.1727

Lavers, D.A., Zsoter, E., Richardson, D.S., Pappenberger, F., 2017. An Assessment of the ECMWF Extreme Forecast Index for Water Vapor Transport during Boreal Winter. Weather Forecast. 32, 1667–1674. https://doi.org/10.1175/WAF-D-17-0073.1

Lee, S.H., Furtado, J.C., Charlton-Perez, A.J., 2019. Wintertime North American Weather Regimes and the Arctic Stratospheric Polar Vortex. Geophys. Res. Lett. 2019GL085592. https://doi.org/10.1029/2019GL085592

Lionello, P., Bhend, J., Buzzi, A., Della-Marta, P.M., Krichak, S.O., Jansà, A., Maheras, P., Sanna, A., Trigo, I.F., Trigo, R., 2006. Chapter 6 Cyclones in the Mediterranean region: Climatology and effects on the environment, in: Lionello, P., Malanotte-Rizzoli, P., Boscolo, R. (Eds.), Developments in Earth and Environmental Sciences, Mediterranean. Elsevier, pp. 325–372. https://doi.org/10.1016/S1571-9197(06)80009-1

Llasat, M.C., Llasat-Botija, M., Petrucci, O., Pasqua, A.A., Rosselló, J., Vinet, F., Boissier, L., 2013. Towards a database on societal impact of Mediterranean floods within the framework of the HYMEX project. Nat. Hazards Earth Syst. Sci. 13, 1337–1350. https://doi.org/10.5194/nhess-13-1337-2013

Llasat, M.C., Llasat-Botija, M., Prat, M.A., Porcú, F., Price, C., Mugnai, A., Lagouvardos, K., Kotroni, V., Katsanos, D., Michaelides, S., Yair, Y., Savvidou, K., Nicolaides, K., 2010. High-impact floods and flash floods in Mediterranean countries: the FLASH preliminary database, in: Advances in Geosciences. Presented at the 10th EGU Plinius Conference on Mediterranean Storms (2008) - 10th Plinius Conference on Mediterranean Storms, Nicosia, Cyprus, 22–24 September 2008, Copernicus GmbH, pp. 47–55. https://doi.org/10.5194/adgeo-23-47-2010

Lolis, C.J., Türkeş, M., 2016. Atmospheric circulation characteristics favouring extreme precipitation in Turkey. Clim. Res. 71, 139–153. https://doi.org/10.3354/cr01433

Mariotti, A., Struglia, M.V., Zeng, N., Lau, K.-M., 2002. The Hydrological Cycle in the Mediterranean Region and Implications for the Water Budget of the Mediterranean Sea. J. Clim. 15, 1674–1690. https://doi.org/10.1175/1520-0442(2002)015<1674:THCITM>2.0.CO;2

Martin-Vide, J., Lopez-Bustins, J.-A., 2006. The Western Mediterranean Oscillation and rainfall in the Iberian Peninsula. Int. J. Climatol. 26, 1455–1475. https://doi.org/10.1002/joc.1388

Merino, A., Fernández-Vaquero, M., López, L., Fernández-González, S., Hermida, L., Sánchez, J.L., García-Ortega, E., Gascón, E., 2016. Large-scale patterns of daily precipitation extremes on the Iberian Peninsula: PRECIPITATION EXTREMES ON THE IBERIAN PENINSULA. Int. J. Climatol. 36, 3873–3891. https://doi.org/10.1002/joc.4601

Neal, R., Fereday, D., Crocker, R., Comer, R.E., 2016. A flexible approach to defining weather patterns and their application in weather forecasting over Europe. Meteorol. Appl. 23, 389–400. https://doi.org/10.1002/met.1563

Olmo, M., Bettolli, M.L., Rusticucci, M., 2020. Atmospheric circulation influence on temperature and precipitation individual and compound daily extreme events: Spatial variability and trends over southern South America. Weather Clim. Extrem. 100267. https://doi.org/10.1016/j.wace.2020.100267

Papalexiou, S.M., Montanari, A., 2019. Global and Regional Increase of Precipitation Extremes Under Global Warming. Water Resour. Res. 55, 4901–4914. https://doi.org/10.1029/2018WR024067

Pavan, V., Antolini, G., Barbiero, R., Berni, N., Brunier, F., Cacciamani, C., Cagnati, A., Cazzuli, O., Cicogna, A., De Luigi, C., Di Carlo, E., Francioni, M., Maraldo, L., Marigo, G., Micheletti, S., Onorato, L., Panettieri, E., Pellegrini, U., Pelosini, R., Piccinini, D., Ratto, S., Ronchi, C., Rusca, L., Sofia, S., Stelluti, M., Tomozeiu, R., Torrigiani Malaspina, T., 2019. High resolution climate precipitation analysis for north-central Italy, 1961–2015. Clim. Dyn. 52, 3435–3453. https://doi.org/10.1007/s00382-018-4337-6

Pedregosa, F., Varoquaux, G., Gramfort, A., Michel, V., Thirion, B., Grisel, O., Blondel, M., Prettenhofer, P., Weiss, R., Dubourg, V., Vanderplas, J., Passos, A., Cournapeau, D., Brucher, M., Perrot, M., Duchesnay, É., 2011. Scikit-learn: Machine Learning in Python. J. Mach. Learn. Res. 12, 2825–2830.

Pfahl, S., 2014. Characterising the relationship between weather extremes in Europe and synoptic circulation features. Nat. Hazards Earth Syst. Sci. 14, 1461–1475. https://doi.org/10.5194/nhess-14-1461-2014

Quadrelli, R., Pavan, V., Molteni, F., 2001. Wintertime variability of Mediterranean precipitation and its links with large-scale circulation anomalies. Clim. Dyn. 17, 457–466. https://doi.org/10.1007/s003820000121

Raveh-Rubin, S., Wernli, H., 2015. Large-scale wind and precipitation extremes in the Mediterranean: a climatological analysis for 1979–2012. Q. J. R. Meteorol. Soc. 141, 2404–2417. https://doi.org/10.1002/qj.2531

Rigo, T., Berenguer, M., Llasat, M. del C., 2019. An improved analysis of mesoscale convective systems in the western Mediterranean using weather radar. Atmospheric Res. 227, 147–156. https://doi.org/10.1016/j.atmosres.2019.05.001

Romero, R., Doswell, C.A., Riosalido, R., 2001. Observations and Fine-Grid Simulations of a Convective Outbreak in Northeastern Spain: Importance of Diurnal Forcing and Convective Cold Pools. Mon. Weather Rev. 129, 2157–2182. https://doi.org/10.1175/1520-0493(2001)129<2157:OAFGSO>2.0.CO;2

Toreti, A., Giannakaki, P., Martius, O., 2016. Precipitation extremes in the Mediterranean region and associated upper-level synoptic-scale flow structures. Clim. Dyn. 47, 1925–1941. https://doi.org/10.1007/s00382-015-2942-1

Toreti, A., Naveau, P., Zampieri, M., Schindler, A., Scoccimarro, E., Xoplaki, E., Dijkstra, H.A., Gualdi, S., Luterbacher, J., 2013. Projections of global changes in precipitation extremes from Coupled Model Intercomparison Project Phase 5 models. Geophys. Res. Lett. 40, 4887–4892. https://doi.org/10.1002/grl.50940

Toreti, A., Xoplaki, E., Maraun, D., Kuglitsch, F.G., Wanner, H., Luterbacher, J., 2010. Characterisation of extreme winter precipitation in Mediterranean coastal sites and associated anomalous atmospheric circulation patterns. Nat. Hazards Earth Syst. Sci. 10, 1037–1050. https://doi.org/10.5194/nhess-10-1037-2010

Trigo, R., Xoplaki, E., Zorita, E., Luterbacher, J., Krichak, S.O., Alpert, P., Jacobeit, J., Sáenz, J., Fernández, J., González-Rouco, F., Garcia-Herrera, R., Rodo, X., Brunetti, M., Nanni, T., Maugeri, M., Türke§, M., Gimeno, L., Ribera, P., Brunet, M., Trigo, I.F., Crepon, M., Mariotti, A., 2006. Chapter 3 Relations between variability in the Mediterranean region and mid-latitude variability, in: Lionello, P., Malanotte-Rizzoli, P., Boscolo, R. (Eds.), Developments in Earth and Environmental Sciences, Mediterranean. Elsevier, pp. 179–226. https://doi.org/10.1016/S1571-9197(06)80006-6

Türkeş, M., Erlat, E., 2005. Climatological responses of winter precipitation in Turkey to variability of the North Atlantic Oscillation during the period 1930–2001. Theor. Appl. Climatol. 81, 45–69. https://doi.org/10.1007/s00704-004-0084-1

Vautard, R., 1990. Multiple Weather Regimes over the North Atlantic: Analysis of Precursors and Successors. Mon. Weather Rev. 118, 2056–2081. https://doi.org/10.1175/1520-0493(1990)118<2056:MWROTN>2.0.CO;2

Vautard, R., Yiou, P., van Oldenborgh, G.-J., Lenderink, G., Thao, S., Ribes, A., Planton, S., Dubuisson, B., Soubeyroux, J.-M., 2015. Extreme Fall 2014 Precipitation in the Cévennes Mountains. Bull. Am. Meteorol. Soc. 96, S56–S60. https://doi.org/10.1175/BAMS-D-15-00088.1

Vicente-Serrano, S.M., Beguería, S., López-Moreno, J.I., El Kenawy, A.M., Angulo-Martínez, M., 2009. Daily atmospheric circulation events and extreme precipitation risk in northeast Spain: Role of the North Atlantic Oscillation, the Western Mediterranean Oscillation, and the Mediterranean Oscillation. J. Geophys. Res. 114, D08106. https://doi.org/10.1029/2008JD011492

Vitart, F., 2014. Evolution of ECMWF sub-seasonal forecast skill scores. Q. J. R. Meteorol. Soc. 140, 1889–1899. https://doi.org/10.1002/qj.2256

Walker, G.T., Bliss, E.W., 1932. World Weather V. Mem. R. Meteorol. Soc., Memoirs of the Royal Meteorological Society 4, 53–84.

Wilks, D.S., 2011. Statistical Methods in the Atmospheric Sciences. Academic Press.

Xoplaki, E., González-Rouco, J.F., Luterbacher, J., Wanner, H., 2004. Wet season Mediterranean precipitation variability: influence of large-scale dynamics and trends. Clim. Dyn. 23, 63–78. https://doi.org/10.1007/s00382-004-0422-0

Xoplaki, E., Trigo, R.M., García-Herrera, R., Barriopedro, D., D'Andrea, F., Fischer, E.M., Gimeno, L., Gouveia, C., Hernández, E., Kuglitsch, F.G., Mariotti, A., Nieto, R., Pinto, J.G., Pozo-Vázquez, D., Saaroni, H., Toreti, A., Trigo, I.F., Vicente-Serrano, S.M., Yiou, P., Ziv, B., 2012. Large-Scale Atmospheric Circulation Driving Extreme Climate Events in the Mediterranean and its Related Impacts, in: Lionello, P. (Ed.), The Climate of the Mediterranean Region. Elsevier, Oxford, pp. 347–417. https://doi.org/10.1016/B978-0-12-416042-2.00006-9

Yiou, P., Nogaj, M., 2004. Extreme climatic events and weather regimes over the North Atlantic: When and where?: WEATHER REGIMES AND EXTREMES. Geophys. Res. Lett. 31, n/a-n/a. https://doi.org/10.1029/2003GL019119

## Supplementary figures

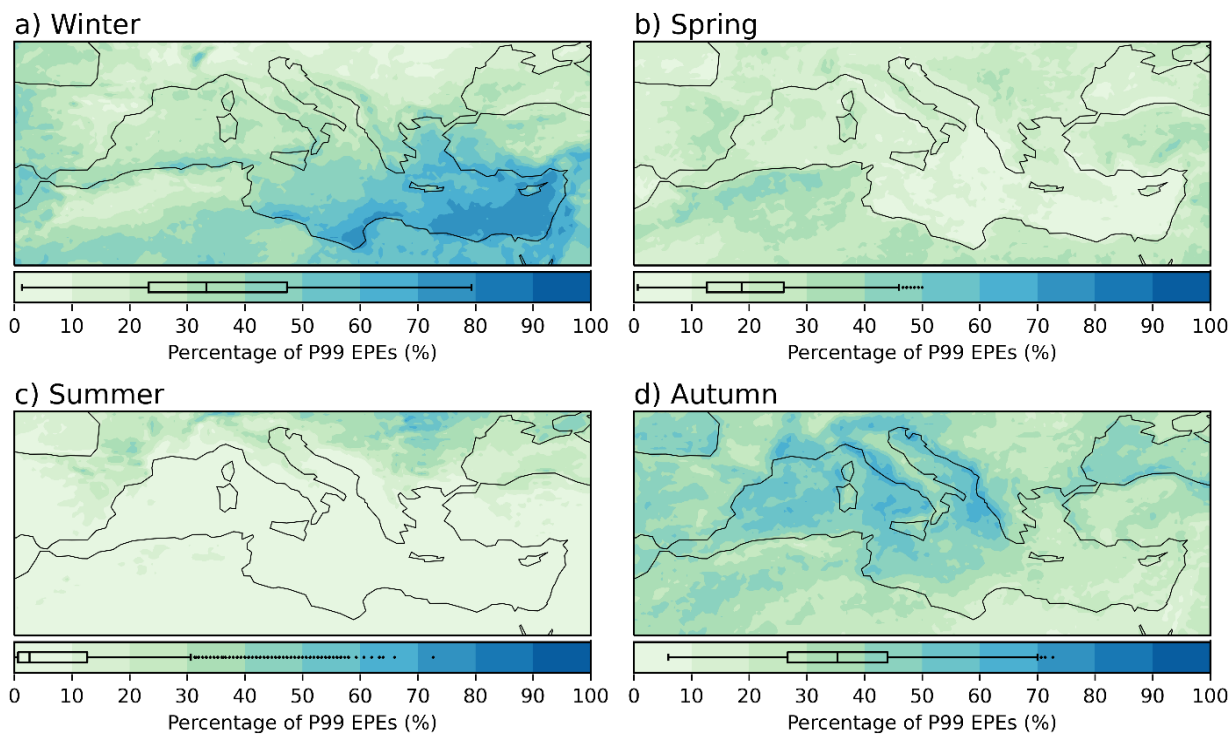

*Figure S1 Percentage of P99 EPEs per season. The boxplots are as in Figure 1.*

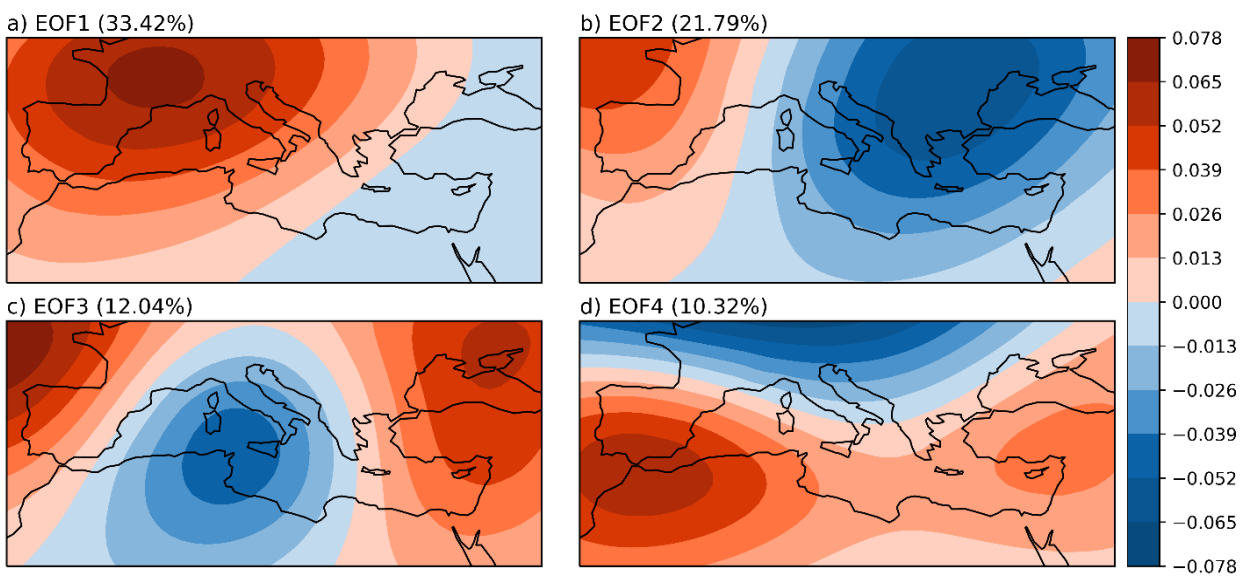

*Figure S2 The first 4 EOFs for Z500 anomalies. The color shades represent the loading scores of each grid cell to the EOFs. The scores are scaled so that the sum of their squares equals to 1. The numbers in the parenthesis indicate the percentage of total variance that is explained by each component.*

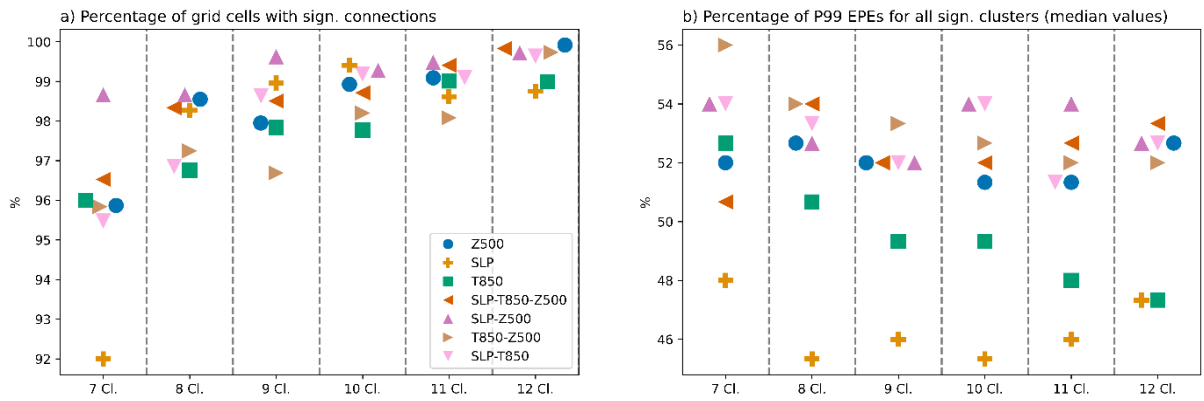

*Figure S3 a) Percentage of grid cells that have statistically significant connection with at least one cluster. b) Percentage of P99 EPEs significantly associated with any of the derived clusters. The results of plot b) represent the median value of all grid cells that have statistically significant connection with at least one cluster.*

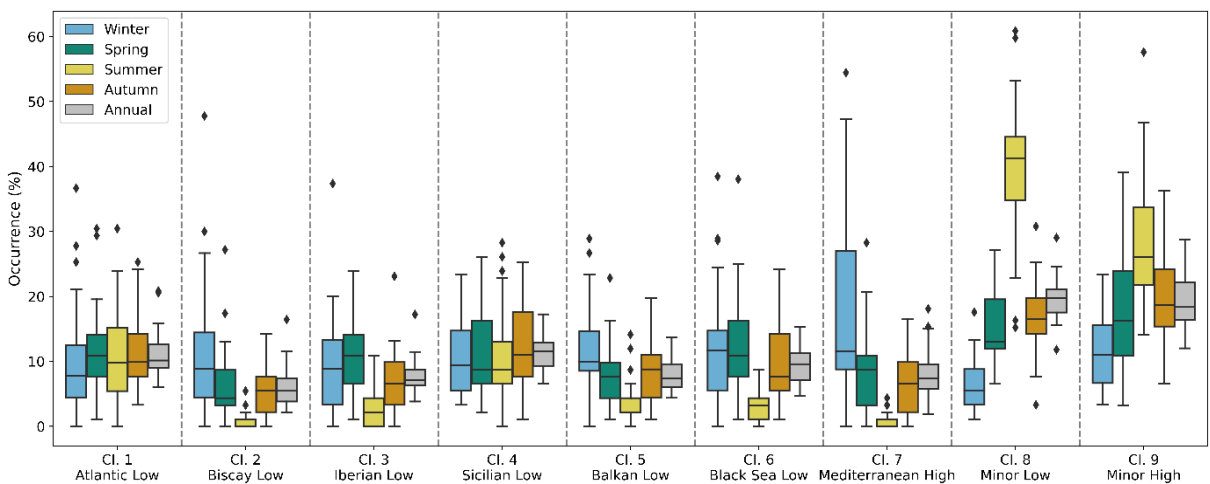

*Figure S4 Seasonal and annual occurrence for the clusters based on SLP and Z500 anomalies. The boxplots are as in Figure 1, based on the available 41 yearly and seasonal data (40 for winter seasons as 1979 winter is not complete).*

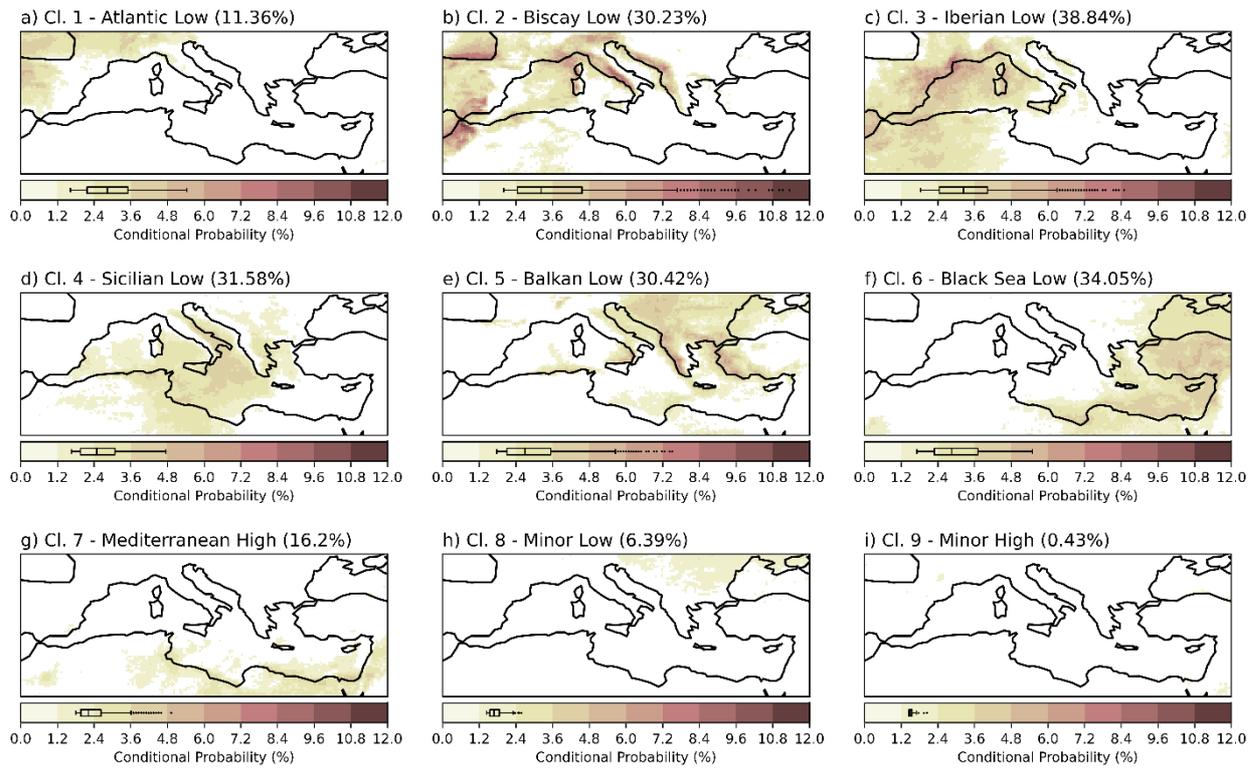

*Figure S5 Connection of weather regimes with P99 EPEs. The conditional probability of EPEs for each weather regime is presented. The results refer only to statistically significant connections, and all other data are masked. The percentages in the parenthesis, refer to the percentage of grid cells that have statistically significant connections with each weather regime. The boxplots are as in Figure 1, but only considering the statistically significant grid cells.*